\documentclass[prd,eqsecnum,twocolumn,showpacs,amsfonts,amssymb]{revtex4}

\usepackage{graphicx}

\usepackage{bm}

\newcommand{\beq}{\begin{equation}}
\newcommand{\eeq}{\end{equation}}
\newcommand{\beqs}{\begin{eqnarray}}
\newcommand{\eeqs}{\end{eqnarray}}
\newcommand{\lsim}{\mathrel{\raisebox{-
.6ex}{$\stackrel{\textstyle<}{\sim}$}}}

\newcommand{\drawsquare}[2]{\hbox{%
\rule{#2pt}{#1pt}\hskip-#2pt
\rule{#1pt}{#2pt}\hskip-#1pt
\rule[#1pt]{#1pt}{#2pt}}\rule[#1pt]{#2pt}{#2pt}\hskip-#2pt
\rule{#2pt}{#1pt}}
\newcommand{\asym}{\raisebox{-3.5pt}{\drawsquare{6.5}{0.4}}\hskip-6.9pt%
        \raisebox{3pt}{\drawsquare{6.5}{0.4}}}

\begin{document}

\title{Ultraviolet Extension of a Model with Dynamical Electroweak 
Symmetry Breaking by Both Top-Quark and Technifermion Condensates} 

\author{Thomas A. Ryttov}

\author{Robert Shrock}

\affiliation{
C. N. Yang Institute for Theoretical Physics \\
Stony Brook University \\
Stony Brook, NY 11794}

\begin{abstract}

We construct and analyze an ultraviolet extension of a model in which
electroweak symmetry breaking is due to both technifermion and top-quark
condensates.  The model includes dynamical mechanisms for all of the various
gauge symmetry breakings.  We discuss certain aspects in which it requires
additional ingredients to be more realistic.

\end{abstract}


\maketitle

\section{Introduction} 
\label{intro}

The origin of electroweak symmetry breaking (EWSB) continues to be an
outstanding mystery.  In one class of models this breaking is produced
dynamically by means of an asymptotically free, vectorial, gauge interaction
based on an exact gauge symmetry, commonly called technicolor (TC), that
becomes strongly coupled on the TeV scale, causing the formation of bilinear
technifermion condensates \cite{tc}. In order to communicate the electroweak
symmetry breaking in the technicolor sector to the Standard-Model (SM) fermions
(which are technisinglets), one embeds the technicolor symmetry in a larger
theory called extended technicolor (ETC) \cite{etc}.  A different approach is
based on the idea that because of its large mass, the top quark should play a
special role in electroweak symmetry breaking, and models of this type feature
a top-quark condensate, $\langle \bar t t \rangle$.  An early realization of
this idea made use of (nonrenormalizable) four-fermion operators \cite{nambu},
while later renormalizable models used separate asymptotically free, vectorial
SU(3) gauge interactions acting on the third generation of quarks and on the
first two generations of quarks, denoted as SU(3)$_1$ and SU(3)$_2$,
respectively \cite{hill,tc2}.  These are often called ``topcolor'' models.  In
these models the SU(3)$_1$ interaction becomes sufficiently strong, at a scale
$\Lambda_t$ of order 1 TeV, to produce the $\langle \bar t t \rangle$
condensate.  The SU(3)$_1$ interaction actually treats the $t$ and $b$ quarks
in the same way and hence, by itself, would also produce a $\langle \bar b b
\rangle$ condensate equal, up to small corrections, to $\langle \bar t t
\rangle$, and a resultant dynamical $b$-quark mass essentially equal to $m_t$.
To prevent the formation of such a $\langle \bar b b \rangle$ condensate, these
models include an additional set of hypercharge-type ${\rm U}(1)_1 \otimes {\rm
U}(1)_2$ gauge interactions.  In these models the ${\rm SU}(3)_1 \otimes {\rm
SU}(3)_2$ and ${\rm U}(1)_1 \otimes {\rm U}(1)_2$ symmetries each break to
their respective diagonal subgroups, which are the usual color SU(3)$_c$ and
weak hypercharge U(1)$_Y$ groups.  There has been considerable interest in
hybrid models that combine the properties of technicolor and topcolor and thus
feature both technifermion condensates and a top-quark condensate
\cite{hill}-\cite{cs08}. These are often called ``topcolor-assisted
technicolor'' or TC2 models. In these theories, most of the observed top-quark
mass, $m_t \simeq 173$ GeV, is due to the $\langle \bar t t \rangle$
condensate.  Reviews of these models include \cite{gcvetic}-\cite{sekhar_pdg}.

In this paper we shall carry out an exploratory construction and analysis of an
ultraviolet extension of a TC2 model in which we explicitly specify an
embedding of the TC symmetry in a higher-lying ETC group and dynamical
mechanisms for the necessary breakings of both the ETC group to the TC group
and of the ${\rm SU}(3)_1 \otimes {\rm SU}(3)_2 \otimes {\rm U}(1)_1 \otimes
{\rm U}(1)_2$ group to ${\rm SU}(3)_c \otimes {\rm U}(1)_Y$.  Because our model
does not purport to be complete, we call it an ultraviolet extension rather
than an ultraviolet completion. It is recognized that models that involve a top
quark condensate and associated strong interactions of the top quark at a scale
not too much larger than $m_t$ are tightly constrained by the excellent
agreement between the measured cross section for $p \bar p \to t \bar t X$ at
$\sqrt{s} = 1.96$ TeV from the CDF and D0 experiments at the Fermilab Tevatron
\cite{pdg,cdfttbar,d0ttbar} and perturbative QCD predictions
\cite{ttbartheory}. TC2 models are also subject to bounds from searches for
colorons, top pions, $t \bar t$ resonances, and by constraints from precision
electroweak data \cite{hill}-\cite{hillsimmons}.  Detailed analyses of the
phenomenology of TC2 models have been given in the literature
\cite{tc2}-\cite{cs08}.  Our present work is somewhat complementary to these
analyses, in that we focus on the effort to build an ultraviolet extension of a
TC2 model, although we do comment on some phenomenological implications of this
extension.  Before proceeding, it is appropriate to recall that TC2 models
represent only one among many ideas for physics beyond the Standard Model;
other ideas include, for example, a top-quark seesaw, supersymmetry, and
theories involving higher spacetime dimensions, in particular, ``higgless''
models and string theory. Here we will concentrate on a (four-dimensional) TC2
approach.

This paper is organized as follows.  In Section II we review some necessary
background on TC/ETC and TC2 models.  In Section III we discuss our ultraviolet
extension and analyze its properties.  Section IV contains a brief discussion
of the consequences that would ensue if one tried to build a model including an
${\rm SU}(2)_{L,1} \otimes {\rm SU}(2)_{L,2}$ sector analogous to the ${\rm
SU}(3)_1 \otimes {\rm SU}(3)_2$ and ${\rm U}(1)_1 \otimes {\rm U}(1)_2$
sectors of TC2 theories.  In a concluding section, we summarize the successes
of the model and certain problems that deserve further study.  Some notation
and formulas are contained in an Appendix.

\section{Some Background}

\subsection{TC/ETC Models} 

Here we briefly review some relevant background on models with dynamical
electroweak symmetry breaking, first on TC/ETC models and then on models
featuring top-quark condensates.  Early works on ETC tended to model ETC
effects via four-fermion operators connecting SM fermions and technifermions,
with some assumed values for their coefficients.  More complete studies have
taken on the task of deriving these four-fermion operators by analyses of
renormalizable, reasonably ultraviolet-complete, ETC models.  These models
normally gauge the generational index and combine it with the technicolor
index. Thus, given that the TC and ETC gauge groups are ${\rm SU}(N_{TC})_{TC}
\subset {\rm SU}(N_{ETC})_{ETC}$, one has the relation
\beq
N_{ETC} = N_{gen}+N_{TC} \ , 
\label{netc}
\eeq
where $N_{gen}=3$ denotes the number of oberved SM fermion generations
\cite{ngen3}.  The ETC gauge symmetry breaks in a series of stages, in
one-to-one correspondence with the SM fermion generations, down to the residual
exact technicolor symmetry.  Some recent reviews include
Refs. \cite{hillsimmons}, \cite{nag06}-\cite{sanrev}.  At the highest breaking
scale, denoted $\Lambda_1$, the first-generation fermions split off, and since
they communicate with the EWSB technifermion sector only via ETC gauge bosons
with masses of order $\Lambda_1$, it follows that their masses are the
smallest. The reasonably ultraviolet-complete ETC models of
Refs. \cite{at94}-\cite{kt} used the minimal non-Abelian value $N_{TC}=2$ in
order to reduce technicolor corrections to $W$ and $Z$
propagators. Accordingly, these models employed an SU(5)$_{ETC}$ group.  In
these models, the ETC-breaking scales corresponding to the three generations
exhibit a hierarchy encompassing $\Lambda_1$ of order $10^3$ TeV, an
intermediate scale, $\Lambda_2$, and the smallest scale, $\Lambda_3$ of order a
few TeV.

TC/ETC models in which the SM-nonsinglet fermions transform vectorially under
the ETC gauge group are able to satisfy constraints from flavor-changing
neutral-current processes.  This was shown in Refs. \cite{ckm,kt} to result
from approximate residual generational symmetries.  These models rely upon a
slowly running (walking) TC gauge coupling associated with an approximate
infrared zero of the TC beta function in order to enhance fermion masses
\cite{wtc,chipt}.  They also must rely upon this walking behavior in another
way, namely that in the presence of a slowly running coupling at the TeV scale,
small perturbations by SM gauge interactions have a magnified effect. Thus,
although the SU(3)$_c$ coupling is small at this scale, it provides enhancement
for the condensation of techniquarks, relative to technileptons, and hence
causes the techniquark condensate to occur at a higher scale, with the result
that the dynamically induced techniquark masses are larger than the
technilepton masses.  This, in turn, can explain why the masses of the quarks
are greater than the mass of the charged lepton in each generation. (The very
small masses of neutrinos require a more complicated mechanism, involving a
low-scale seesaw \cite{nt}.)  Furthermore, since the weak hypercharge
interaction favors the condensation of the techniquarks of charge 2/3 in a
one-family model, while inhibiting the condensation of techniquarks of charge
$-1/3$ (see Eqs.  (\ref{yqlyur})-(\ref{yqlydr})), the former naturally condense
at a higher scale then the latter.  This can explain why the charge 2/3 quarks
of the higher two generations are heavier than the charge $-1/3$ quarks
(explaining why $m_u < m_d$ presumably necessitates incorporating effects
of off-diagonal elements of the respective up-quark and down-quark mass
matrices).  However, it is not clear that this effect is large enough to
account for the large mass ratio $m_t/m_b$ without violating custodial-symmetry
constraints and, moreover, is able to produce realistic CKM mixing
\cite{kt,csm}. Indeed, the large value of the top-quark mass was one of the
main motivations for models featuring a $\langle \bar t t \rangle$ condensate.

\subsection{Models with $\langle \bar F F \rangle$ and 
$\langle \bar t t \rangle$ } 

We proceed to discuss some details of TC2 models that will be needed for the
explanation of our ultraviolet extension.  These use a gauge group,
\beq
G_{TC} \otimes G_{ASM} \ , 
\label{gtcasm}
\eeq
where $G_{TC}$ is the technicolor group and $G_{ASM}$ is the augmented SM (ASM)
group
\beqs
G_{ASM} & = & {\rm SU}(3)_1 \otimes {\rm SU}(3)_2 \otimes 
{\rm SU}(2)_L \otimes {\rm U}(1)_1 \otimes {\rm U}(1)_2 \ . 
\cr\cr & & 
\label{gasm}
\eeqs
Our notation for the running gauge couplings (with the scale $\mu$ implicit
here) is $g_{_{TC}}$ and, for the five factor groups in $G_{ASM}$, $g_{c1}$,
$g_{c2}$, $g$, $g'_1$, and $g'_2$.  The running squared couplings are denoted
$\alpha_j \equiv g_j^2/(4\pi)$ for the various factor groups $G_j$. The gauge
symmetry (\ref{gtcasm}) is operative above a scale of order 1 TeV and below the
lowest ETC breaking scale. As discussed above, the SU(3)$_1$ interaction
couples to the third generation of quarks, while the SU(3)$_2$ interaction
couples to the first two generations of quarks.  The SU(3)$_1$ coupling at this
scale is considerably stronger than the SU(3)$_2$ coupling, and, indeed,
becomes strong enough to produce the $\langle \bar t t \rangle$ condensate.

To prevent the formation of a $\langle \bar b b \rangle$ condensate by this
SU(3)$_1$ interaction, TC2 models rely on the ${\rm U}(1)_1 \otimes {\rm
U}(1)_2$ factor group displayed in Eq. (\ref{gasm}).  In early TC2 models, the
U(1)$_1$ and U(1)$_2$ interactions coupled, respectively, to SM fermions of the
third generation, and to SM fermions of the first two generations, according to
their weak hypercharges.  Motivated by constraints from precision electroweak
data, more recent TC2 models \cite{lane05,cs08} have adopted a different set of
${\rm U}(1)_1 \otimes {\rm U}(1)_2$ charge assignments in which the ${\rm
U}(1)_1$ interaction couples in the same manner to all three generations, which
are singlets under U(1)$_2$. These models have thus been characterized as
having flavor-universal hypercharge.  At the scale $\Lambda_t$, the U(1)$_1$ 
interaction is assumed to be strong enough to (i) enhance the
formation of the $\langle \bar t t \rangle = \langle \bar t_L t_R \rangle +
h.c.$ condensate, since the relevant hypercharge product
\beq
(-Y_{Q_L})Y_{u_R} = -\frac{4}{9} 
\label{yqlyur}
\eeq
is attractive, and (ii) prevent the formation of a $\langle \bar b b
\rangle = \langle \bar b_L b_R \rangle + h.c.$ condensate, since the
hypercharge product
\beq
(-Y_{Q_L})Y_{d_R} = + \frac{2}{9} 
\label{yqlydr}
\eeq
is repulsive \cite{qlw}.  However, there are several constraints on the
strength of the U(1)$_1$ coupling.  First, if it were too large, then there
would be excessive violation of custodial symmetry.  Second, since the U(1)$_1$
(as well as U(1)$_2$) gauge interaction is not asymptotically free, a
moderately strong U(1)$_1$ coupling would bring with it the danger of a Landau
pole at an energy not too far above the 1 TeV scale, so that the model could
not be regarded as a self-consistent low-energy effective field theory.  These
constraints have been used in TC2 model-building
\cite{hill}-\cite{hillsimmons}.  Indeed, in view of these constraints and the
fact that, as the energy scale $\mu$ decreases, the U(1)$_1$ coupling gets
weaker while the SU(3)$_1$ coupling gets stronger, there is a rather limited
set of values of couplings and a limited interval in which this scenario can
take place in a self-consistent manner. In particular, the SU(3)$_1$ coupling
at the scale $\Lambda_t$ must be fine-tuned to be only slightly greater than
the critical value for the formation of the $\langle \bar t t \rangle$
condensate, so that a rather weak U(1)$_1$ coupling can still prevent the
formation of a $\langle \bar b b \rangle$ condensate \cite{tc2}-\cite{cs08}.

To the extent that this top-quark mass generation by SU(3)$_1$ is analogous to
the dynamical generation of constituent-quark masses in quantum chromodynamics
(QCD), then, since the latter are of order $\Lambda_{QCD}$, one would infer
that $\Lambda_t$ would be roughly comparable to $m_t$.  More quantitatively,
one can use the approximate relation \cite{psrel}
\beq
f_t^2 \simeq \frac{3 \, \Sigma_t^2}{4\pi^2} \ln \left ( \frac{\Lambda_{int}}
{\Sigma_t} \right ) 
\label{psrel}
\eeq
where $\Lambda_{int}$ represents a cutoff scale characterizing the asymptotic
decay of the dynamical mass $\Sigma_t$, considered as a running quantity.  This
scale, $\Lambda_{int}$, enters in the integral that one calculates in 
deriving this relation.  Setting $\Sigma_t \simeq m_t$ and using the rough
estimate $\Lambda_{int} \simeq 2\Sigma_t$, one obtains $f_t \simeq 70$ GeV.
The technifermion condensation yields an analogous $f_{TC}$.  Both the
top-quark and technifermion condensates transform as $\Delta T_3=1/2$ under
SU(2)$_L$ and $|\Delta Y|=1$ under U(1)$_Y$, and hence produce a $W$ mass
given, to leading order, by
\beq
m_W^2 = \frac{g^2(N_{TD}f_{TC}^2+f_t^2)}{4}
\label{mwsq}
\eeq
where $N_{TD}$ denotes the number of SU(2)$_L$ technidoublets in the theory.
Our ultraviolet extension will use a one-family TC model, so that
$N_{TD}=(N_c+1)=4$.  In the absence of the $f_t^2$ contribution from the
top-quark condensate (i.e., in regular technicolor), Eq. (\ref{mwsq}) would
yield $f_{TC} \simeq 125$ GeV; the value of $f_{TC}$ in the TC2 theory is
slightly reduced by the presence of the $f_t^2$ term.  Since
$f_t^2/(N_{TD}f_{TC}^2) \lsim 0.1$, most of the contribution to the $W$ mass in
the TC2 model is provided by technicolor.  Similar comments apply to the $Z$
mass. 

One should remark on a difference between the generation of dynamical masses
for light quarks in QCD and techniquarks in TC, on the one hand, and the
generation of the top-quark mass in TC2 theories, on the other hand.  In the
former two cases, the gauge interactions responsible for the condensates and
resultant dynamical fermion masses are exact.  In contrast, SU(3)$_1$ is broken
at a scale comparable to the scale $\Lambda_t$ where it gets strong and
produces the $\langle \bar t t \rangle$ condensate.  It is thus plausible that,
to compensate for this, $\Lambda_t$ should be somewhat larger than $\Sigma_t
\simeq m_t$, say of order 1 TeV, and we will assume this approximate value
here. Slightly below the $\Lambda_t$ scale, the ${\rm SU}(3)_1 \otimes {\rm
SU}(3)_2$ symmetry group breaks to its diagonal subgroup that treats all
generations symmetrically, namely usual color SU(3)$_c$, while the ${\rm
U}(1)_1 \otimes {\rm U}(1)_2$ symmetry group breaks to a diagonal subgroup,
which is the usual weak hypercharge, U(1)$_Y$.  Thus, one has the symmetry
breaking $G_{ASM} \to G_{SM}$, where $G_{SM} = {\rm SU}(3)_c \otimes {\rm
SU}(2)_L \otimes {\rm U}(1)_Y$.

We shall use the more modern type of TC2 model with a U(1)$_1$ coupling
universally to all generations \cite{lane05,cs08} to serve as the basis for our
ultraviolet extension.  We display SM fermion representations in this type of
model below.  In our notation, the three numbers in parentheses are the
dimensions of the representations of the three non-Abelian factor groups in
$G_{ASM}$; the subscripts are the U(1)$_1$ and U(1)$_2$ hypercharges; and $a$
and $a' \in \{1,2,3\}$ are SU(3)$_1$ and SU(3)$_2$ indices:
\beq
 {u^{a'} \choose d^{a'}}_L, \ {c^{a'} \choose s^{a'}}_L  \ : 
\quad 2(1,3,2)_{1/3,0}
\label{ql}
\eeq
\beq
{t^{a} \choose b^{a}}_L \ : \quad (3,1,2)_{1/3,0}
\label{tbl}
\eeq
\beq
u^{a'}_R, \ c^{a'}_R  \ :  2(1,3,1)_{4/3,0}; \quad t^{a}_R \ : (3,1,1)_{4/3,0}
\label{uctr}
\eeq
\beq
 d^{a'}_R, \ s^{a'}_R  \ : 2(1,3,1)_{-2/3,0}; \quad b^{a}_R \ : (3,1,1)_{-2/3,0}
\label{dsbr}
\eeq
\beq
{\nu_e \choose e}_L, \ {\nu_\mu \choose \mu}_L  \ : \quad 2(1,1,2)_{-1,0}
\label{emul}
\eeq
\beq
{\nu_\tau \choose \tau}_L \ : \quad (1,1,2)_{-1,0}
\label{taul}
\eeq
and
\beq
 e_R, \ \mu_R \ :  2(1,1,1)_{-2,0}; \quad \tau_R \ : (1,1,1)_{-2,0} \ . 
\label{emtr}
\eeq
It is an option whether one explicitly includes right-handed
electroweak-singlet neutrinos, since they are singlets under $G_{ASM}$.

\section{Construction and Assessment of an Ultraviolet Extension} 

\subsection{General Structure}

We find that it is not possible to have the usual ETC structure given by
Eq. (\ref{netc}).  The reason is quite fundamental; the full ETC symmetry is
incompatible with the essential feature of the model, namely the fact that the
first two generations of SM fermions transform according to different
representations of $G_{ASM}$ than the third generation.  In other words, 
the ETC symmetry implies, {\it a fortiori}, that unitary transformations that
mix up the three left-handed SU(2)$_L$ quark doublets leave the theory
invariant, but this requirement is incompatible with the assignment of the
first two generations of these quark doublets to the representation
$(1,3)$ of ${\rm SU}(3)_1 \otimes {\rm SU}(3)_2$ and the third to the 
different representation of this group, $(3,1)$. 
Similarly, the ETC symmetry implies, {\it a fortiori}, that
unitary transformations that mix up the three generations of 
right-handed up-type quarks and, separately, the three generations of down-type
quarks leave the theory invariant, but this requirement is incompatible with 
the assignment of the first two generations of these quark fields to the 
representation $(1,3)$ of ${\rm SU}(3)_1 \otimes {\rm SU}(3)_2$ and the third
generation to the $(3,1)$ representation.  

In view of this fundamental incompatibility, we shall construct the ETC group
by embedding the first two generations of SM fermions together with the
technifermions in ETC multiplets.  Hence, for our ETC model, the relation
(\ref{netc}) is altered to read
\beq
N_{ETC} = N_{gen}-1+N_{TC} = 2+N_{TC} \ . 
\label{netcprime}
\eeq
As before, in order to minimize TC corrections to $W$ and $Z$ propagators, we
again choose the minimal non-Abelian value, $N_{TC}=2$, so our ETC group is
SU(4)$_{ETC}$.  We thus consider a model that, at a high scale, is invariant
under the gauge symmetry
\beq
G = G_{ETCA} \otimes G_{ASM} \ , 
\label{ggen}
\eeq
where
\beq
G_{ETCA} = {\rm SU}(4)_{ETC} \otimes {\rm SU}(2)_{HC} \otimes {\rm SU}(3)_{MC} 
\otimes {\rm SU}(2)_{UC} 
\label{getca}
\eeq
and $G_{ASM}$ was given in Eq. (\ref{gasm}).  The group $G_{ETCA}$ contains the
ETC group, SU(4)$_{ETC}$, together with three additional gauge interactions
(the subscript $A$ in $ETCA$ refers to these \underline{a}dditional
interactions): (i) hypercolor (HC) SU(2)$_{HC}$, which helps in the breaking of
SU(4)$_{ETC}$ in two sequential stages, to SU(3)$_{ETC}$ and then to the
residual exact technicolor group, SU(2)$_{TC}$; (ii) metacolor (MC)
SU(3)$_{MC}$, which breaks ${\rm SU}(3)_1 \otimes {\rm SU}(3)_2$ to the
diagonal subgroup, color SU(3)$_c$; and (iii) ultracolor (UC) SU(2)$_{UC}$,
which breaks ${\rm U}(1)_1 \otimes {\rm U}(1)_2$ to the diagonal subgroup, weak
hypercharge U(1)$_Y$.  With the fermion content to be delineated below, all of
the four gauge interactions in $G_{ETCA}$ are asymptotically free.

The fermions with SM quantum numbers, including the usual SM fermions and the
technifermions, are assigned to the representations displayed below. We use
notation such that the four numbers in the parentheses are the dimensions
of the representations of the group 
\beq
{\rm SU}(4)_{ETC} \otimes {\rm SU}(3)_1 \otimes {\rm SU}(3)_2 \otimes 
{\rm SU}(2)_L \ , 
\label{ag}
\eeq
while the two subscripts are the hypercharges for the gauge groups U(1)$_1$ and
U(1)$_2$, respectively. Since all of these fermions are singlets under the
additional gauge interactions in $G_{ASM}$, namely ${\rm SU}(2)_{HC} \otimes
{\rm SU}(3)_{MC} \otimes {\rm SU}(2)_{UC}$, we do not include these factor
groups in the listings.  The index $i$ is an SU(4)$_{ETC}$ index, with $i=1,2$
referring to the first two generations and $i=3,4$ being SU(2)$_{TC}$ gauge
indices.  As before, e.g. in \cite{ckm}, we use a compact notation in which
$u^{a',1} \equiv u^{a'}$, $u^{a',2} \equiv c^{a'}$, $d^{a',1} \equiv d^{a'}$,
$d^{a',2} \equiv s^{a'}$, $e^1 \equiv e$, and $e^2 \equiv \mu$. The fermion
representations are
\beq
Q^{a',i}_L = \left( \begin{array}{cccc}
    u^{a',1} & u^{a',2} & u^{a',3} & u^{a',4} \\
    d^{a',1} & d^{a',2} & d^{a',3} & d^{a',4} \end{array} \right )_L \ : \quad
(4,1,3,2)_{1/3,0}
\label{qlgen}
\eeq
\beq
{t^{a} \choose b^{a}_L}_L \ : \quad (1,3,1,2)_{1/3,0}
\label{tblgen}
\eeq
\beq
(u^{a',1},u^{a',2},u^{a',3},u^{a',4})_R \ : \quad (4,1,3,1)_{4/3,0}
\label{urgen}
\eeq
\beq
 \quad t^{a}_R \ : (1,3,1,1)_{4/3,0}
\label{trgen}
\eeq
\beq
(d^{a',1},d^{a',2},d^{a',3},d^{a',4})_R \ :  \quad (4,1,3,1)_{-2/3,0}
\label{drgen}
\eeq
\beq
 \quad b^{a}_R \ : \quad (1,3,1,1)_{-2/3,0}
\label{brgen}
\eeq
\beq
L_L = \left( \begin{array}{cccc}
    \nu^1 & \nu^2 & \nu^3 & \nu^4 \\
    e^1  & e^2   & e^3   & e^4 \end{array} \right )_L \ : \quad
(4,1,1,2)_{-1,0}
\label{elgen}
\eeq
\beq
{\nu_\tau \choose \tau}_L \ : \quad (1,1,1,2)_{-1,0}
\label{taulgen}
\eeq
\beq
(e^1,e^2,e^3,e^4)_R \ ; \quad (4,1,1,1)_{-2,0}
\label{ergen}
\eeq
and
\beq
\tau_R \ : \quad (1,1,1,1)_{-2,0} \ . 
\label{taur}
\eeq
As was alluded to above, this is thus a one-family technicolor model
\cite{ttf}.  Given the motivation for the structure of the model, it is clear
why there are no SM fermions that are simultaneously nonsinglets under both
SU(3)$_1$ and SU(3)$_2$.  As pointed out above, one cannot embed all of the
generations of each type of fermion in a single corresponding ETC multiplet,
since there is an incompatibility between the essential ETC feature of treating
the three generations in a symmetric manner at the high scale and the fact that
in these types of models the first two generations are subject to different
gauge symmetries than the third generation. Hence, with the present embedding
in SU(4)$_{ETC}$, there is only mixing of the first two generations with each
other, but no full three-generation CKM (Cabibbo-Kobayashi-Maskawa) mixing.
Thus, $V_{ub}=V_{cb}=V_{td}=V_{ts}=0$. (Indeed, the observed CKM quark mixing
represents the difference in mixings between the up-quark and down-quark
sectors, so three-generational mixings in these individual sectors are
necessary but not sufficient to fit the observed CKM mixing.)  The fact that
the third-generation quarks transform differently under ${\rm SU}(3)_1 \otimes
{\rm SU}(3)_2$ than the first two generations of quarks was recognized in early
TC2 model-building to pose a challenge to getting full CKM mixing
\cite{hill,tc2}, and this problem manifests itself directly in our UV
extension.  This shows that further ingredients are required for a satisfactory
larger ultraviolet completion.

\subsection{Generalities on Fermion Condensation Channels}

In general, in an asymptotically free gauge theory involving possible
condensation of fermions transforming according to the representations $R_1$
and $R_2$ of the gauge group $G_j$ to a condensate transforming as $R_{cond}$,
an approximate measure of attractiveness of this channel 
\beq
R_1 \times R_2 \to R_{cond} 
\label{rchannel}
\eeq
is
\beq
\Delta C_2 = C_2(R_1) + C_2(R_2) - C_2(R_{cond}) \ , 
\label{deltac2}
\eeq
where $C_2(R)$ is the quadratic Casimir invariant for the representation $R$
\cite{casimir}.  If several possible condensation channels are possible, it is
expected that condensation occurs in the most attractive channel (MAC), i.e.,
the one with the largest value of $\Delta C_2$.  For a vectorial gauge
interaction, the most attractive channel is $R \times \bar R \to 1$, producing
a condensate in the singlet representation of the gauge group (with $\Delta C_2
= 2C_2(R)$) and thus preserving the gauge invariance.  In this case, as the
reference energy scale $\mu$ decreases from large values where the gauge
interaction is weak, this condensation is expected to occur when $\alpha_j(\mu)
C_2(R)$ exceeds a value of order unity.  Some results relevant to this are
given in the Appendix.  For a particular asymptotically free gauge interaction,
one must check to see whether, given its (light or massless) fermion content,
it will evolve from high scales where it is weakly coupled to lower scales in a
manner that leads to a growth in the coupling that is sufficient to trigger
fermion condensation, or whether, alternatively, its coupling could approach an
infrared fixed point that is too small for such condensation to occur.  In the
latter case, this gauge interaction would not spontaneously break its chiral
symmetries.  In the model studied here, it is required that the SU(2)$_{HC}$,
SU(3)$_1$, SU(3)$_{MC}$, SU(2)$_{UC}$, and SU(2)$_{TC}$ gauge interactions
produce various condensates, and we will show that this is, indeed, consistent
with the sets of nonsinglet fermions subject to these respective interactions.
Throughout our analysis, it is understood that there are theoretical
uncertainties inherent in analyzing such strong-coupling phenomena as fermion
condensation.

We proceed to describe the fermion contents of the rest of the model. Since 
the full model is a chiral gauge theory, it follows that the
Lagrangian describing the physics at a high scale $\sim 10^4$ TeV has no
fermion mass terms.  An analysis of global symmetries is of interest especially
since some of these symmetries are broken by condensates produced by gauge
symmetries that become strong at various lower energy scales. We shall discuss
these global symmetries below.

\subsection{SU(2)$_{HC}$ Sector}

We shall need a set of fermions whose role is to break the SU(4)$_{ETC}$
symmetry in two stages down to SU(2)$_{TC}$.  These fermions are singlets under
all of the gauge symmetries except ${\rm SU}(4)_{ETC} \otimes {\rm
SU}(2)_{HC}$, so we only list their dimensionalities under these two groups:
\beq
\psi_{j,R} \ : \quad (\bar 4,1)
\label{ngen2_psi}
\eeq
\beq
\chi^{j,\alpha}_R \ : \quad (4,2)
\label{ngen2_chi}
\eeq
and
\beq
\zeta^{jk,\alpha}_R \ : \quad (6,2) \ , 
\label{ngen2_zeta}
\eeq
where $j$ and $\alpha$ are SU(4)$_{ETC}$ and SU(2)$_{HC}$ indices,
respectively.  The 6-dimensional representation of SU(4) is the antisymmetric
rank-2 tensor representation, $\asym$, which is self-conjugate (and hence has
zero SU(4)$_{ETC}$ gauge anomaly).

We next discuss the global flavor symmetries involving hypercolor-nonsinglet
fermions.  The fact that the $\chi^{j,\alpha}_R$ and $\zeta^{jk,\alpha}_R$
fields are nonsinglets under two interactions, namely SU(4)$_{ETC}$ and
SU(2)$_{HC}$, that become strongly coupled at comparable scales ($\Lambda_1
\sim \Lambda_{HC} \sim 10^3$ TeV) plays an important role in the determination
of this global chiral symmetry. In the hypothetical limit where, at a given
scale, the SU(2)$_{HC}$ coupling were imagined to be much stronger than the
SU(4)$_{ETC}$ coupling, it would follow that the sector of HC-nonsinglet
fermions would be invariant under the classical global flavor symmetry group
${\rm U}(4)_\chi \otimes {\rm U}(6)_\zeta$, or equivalently, ${\rm SU}(4)_\chi
\otimes {\rm SU}(6)_\zeta \otimes {\rm U}(1)_\chi \otimes {\rm U}(1)_\zeta$.
Here, the global U(1)$_\chi$ and U(1)$_\zeta$ transformations are defined to
rephase $\chi^{j,\alpha}_R$ and $\zeta^{jk,\alpha}_R$, respectively.  Both of
these global U(1) symmetries are broken by SU(2)$_{HC}$ instantons, with one
linear combination remaining unbroken.  Just as SU(2)$_L$ instantons break
quark number, $N_q$, and lepton number, $N_L \equiv L$, but preserve
$N_cN_q-N_L = B-L$, so also the SU(2)$_{HC}$ instantons preserve the linear
combination $4N_\chi - 6N_\zeta$, or equivalently, $2N_\chi-3N_\zeta$.  Let us
denote the corresponding global ($g$) number symmetry as U(1)$_{HCg}$.  Thus,
if SU(4)$_{ETC}$ interactions could be neglected relative to SU(2)$_{HC}$, then
the actual global chiral symmetry group of this sector would be ${\rm
SU}(4)_\chi \otimes {\rm SU}(6)_\zeta \otimes {\rm U}(1)_{HCg}$.  However,
although the SU(2)$_{HC}$ interaction is stronger than the SU(4)$_{ETC}$
interaction, the latter is never negligible, and hence the global flavor
symmetry group is not ${\rm SU}(4)_\chi \otimes {\rm SU}(6)_\zeta \otimes 
{\rm U}(1)_{HCg}$ symmetry, but instead only U(1)$_{HCg}$.

\subsection{SU(3)$_{MC}$ Sector} 

The fermions in the second set are involved with the breaking of ${\rm
SU}(3)_1 \otimes {\rm SU}(3)_2$ to the diagonal color subgroup, color
SU(3)$_c$.  This set contains nonsinglets under only the group
${\rm SU}(3)_{MC} \otimes {\rm SU}(3)_1 \otimes {\rm SU}(3)_2$ (i.e., they all
have zero U(1)$_1$ and U(2)$_2$ hypercharges); 
with respect to this group, the fermions transform as 
\beq
\xi^{a,\lambda}_R \ : \quad (3,3,1)
\label{xiR}
\eeq
\beq
\eta^{a}_{p,L} \ : \quad (1,3,1) \ , \quad p=1,2,3
\label{etaL}
\eeq
\beq
\xi^{a',\lambda}_L \ : \quad (3,1,3)
\label{xiL}
\eeq
and
\beq
\eta^{a'}_{p,R} \ : \quad (1,1,3) \ , \quad p=1,2,3 \ , 
\label{etaR}
\eeq
where $\lambda$, $a$, and $a'$ are SU(3)$_{MC}$, SU(3)$_1$, and SU(3)$_2$ gauge
indices.  (This set of fermion fields could be written equivalently in
holomorphic form as all right-handed or all left-handed fields by using
appropriate complex conjugates.)

The fermions that are nonsinglets under SU(3)$_1$ include the third-generation
quarks and the fermions $\xi^{a,\lambda}_R$ and $\eta^a_{p,L}$ in Eqs.
(\ref{xiR}) and (\ref{etaL}).  In order to determine the operative global
flavor symmetry involving MC-nonsinglet fermions slightly above 1 TeV, one must
take account of the fact that both the SU(3)$_1$ and SU(3)$_{MC}$ interactions
become strongly coupled at this scale.  In accordance with constraints from
custodial symmetry, we shall assume that the U(1)$_1$ gauge coupling is
sufficiently weakly coupled so that it, together with the other
(non-technicolor) gauge interactions can be neglected, in a leading
approximation, in considering the global flavor symmetry. Then the classical
global flavor symmetry group at this scale would be
\beqs
& & {\rm U}(2)_{(t,b)_L} \otimes {\rm U}(2)_{(t,b)_R} \otimes 
{\rm U}(1)_{\xi_R} \otimes {\rm U}(3)_{\eta_L} \otimes {\rm U}(3)_{\xi_L} \ ,
\cr\cr
& & 
\label{su31cgsym}
\eeqs
or equivalently, 
\begin{widetext}
\beq
{\rm SU}(2)_{(t,b)_L} \otimes {\rm SU}(2)_{(t,b)_R} \otimes 
{\rm U}(1)_{(t,b)_V} \otimes {\rm U}(1)_{(t,b)_A} \otimes 
{\rm U}(1)_{\xi_R} \otimes {\rm SU}(3)_{\eta_L} \otimes {\rm U}(1)_{\eta_L} 
\otimes {\rm SU}(3)_{\xi_L} \otimes {\rm U}(1)_{\xi_L} \ . 
\label{su31cgsymexplicit}
\eeq
\end{widetext}
Here, SU(2)$_{(t,b)_L}$ and SU(2)$_{(t,b)_R}$ operate on the left- and
right-handed chiral $t$ and $b$ fields; U(1)$_{(t,b)_V}$ and U(1)$_{(t,b)_A}$
are vector and axial-vector U(1)'s operating on $t$ and $b$; the U(1)$_{\xi_R}$
rephases the $\xi^{a,\lambda}_R$ fields (for fixed $a$ and $\lambda$); the
U(3)$_{\eta_L}$ operates on the three $\eta^a_{p,L}$ fields (with $a$ fixed,
and $p=1,2,3$); and the U(3)$_{\xi_L}$ operates on the three
$\xi^{a',\lambda}_L$ fields (with $\lambda$ fixed and $a'=1,2,3$).  SU(3)$_1$
instantons leave U(1)$_{(t,b)_V}$ invariant but break U(1)$_{(t,b)_A}$,
U(1)$_{\xi_R}$, and U(1)$_{\eta_L}$.  SU(3)$_{MC}$ instantons break the
U(1)$_{\xi_R}$ and U(1)$_{\xi_L}$ symmetries.  (Thus, in particular, the
U(1)$_{\xi_R}$ symmetry is broken by both SU(3)$_1$ and SU(3)$_{MC}$
instantons.)  From the broken U(1)$_{(t,b)_A}$ and U(1)$_{\eta_L}$ symmetries
one can form a linear combination, which we denote U(1)$'$, that is preserved
by the SU(3)$_1$ instantons.  To form a conserved axial-vector current
involving the $\xi_R$ field, one needs to cancel the divergences due to both
the SU(3)$_1$ and SU(3)$_{MC}$ instantons, which requires a linear combination
of the axial-vector currents involving the $\eta_L$ and $\xi_L$,
respectively. We denote this conserved global symmetry as U(1)$_{scm}$ (where
$scm=$ ``strongly coupled, mixed'').  Thus, with the above-mentioned provisos
that other gauge interactions can be considered negligible, the actual quantum
global flavor symmetry involving fermions that are nonsinglets under the
strongly coupled SU(3)$_1$ and SU(3)$_{MC}$ gauge symmetries is
\begin{widetext}
\beq 
{\rm SU}(2)_{(t,b)_L} \otimes {\rm SU}(2)_{(t,b)_R} \otimes {\rm
U}(1)_{(t,b)_V} \otimes {\rm SU}(3)_{\eta_L} \otimes {\rm U}(1)' \otimes
{\rm SU}(3)_{\xi_L} \otimes {\rm U}(1)_{scm} \ .  
\label{su31globalsym}
\eeq
\end{widetext}

\subsection{SU(2)$_{UC}$ Sector}

The third set of fermions is involved with the breaking of ${\rm U}(1)_1
\otimes {\rm U}(1)_2$ to the diagonal subgroup, weak hypercharge U(1)$_Y$.
This set contains nonsinglets under only the group
${\rm SU}(2)_{UC} \otimes {\rm U}(1)_1 \otimes {\rm U}(1)_2$,  
and, with respect to this group, the fields transform as 
\beq
\Omega^{\hat{\alpha}}_R \ : \quad 2_{y,0}
\label{Omega}
\eeq
\beq
\omega_{p,R} \ : \quad 1_{-y,0} \ , \quad p=1,2
\label{omega}
\eeq
\beq
\tilde \Omega^{\hat{\alpha}}_R \ : \quad 2_{0,-y}
\label{Omegatilde}
\eeq
and
\beq
\tilde \omega_{p,R} \ : \quad 1_{0,y} \ , \quad p=1,2 \ , 
\label{omegatilde}
\eeq
where $\hat{\alpha}$ is an SU(2)$_{UC}$ gauge index, the subscripts denote the
hypercharges with respect to U(1)$_1$ and U(1)$_2$, and $y \ne 0$.  This
SU(2)$_{UC}$ sector has a classical global ${\rm U}(1)_\Omega \otimes {\rm
U}(1)_{\hat\Omega}$ symmetry, where U(1)$_{\hat \Omega}$ and
U(1)$_{\tilde{\hat\Omega}}$ rephase the $\Omega^{\hat{\alpha}}_R$ and $\tilde
\Omega^{\hat{\alpha}}_R$ fields, respectively.  Both of these U(1)'s are
broken by the SU(2)$_{UC}$ instantons, but the combination corresponding to
$N_{UCg} = N_\Omega - N_{\hat\Omega}$ is preserved.  We denote this as
U(1)$_{UCg}$.

As is evident from these fermion representation assignments, we have chosen to
construct the ultraviolet extension to have a modular structure, in which one
sector is responsible for the breaking of ${\rm SU}(3)_1 \otimes {\rm SU}(3)_2$
to SU(3)$_c$ and another is responsible for the breaking of ${\rm U}(1)_1
\otimes {\rm U}(1)_2$ to U(1)$_Y$, rather than trying to accomplish this
breaking with a single sector.  While this makes the model somewhat
complicated, it actually simplifies some aspects of the analysis, such as
checking anomaly cancellation and determining condensation channels.  One could
also investigate models in which one tries to use a single sector to carry out
both of these symmetry breakings.

\subsection{Anomaly Cancellation}

Since the full model is a chiral gauge theory, it is necessary to check that it
is free of any gauge or global anomalies. Given the modular construction of the
theory, we can divide the analysis of anomalies into several parts.  The first
involves contributions of fermions that are nonsinglets under the SU(4)$_{ETC}$
group. We first observe that the SU(4)$_{ETC}$ anomalies of the $\psi_{i,R}$
and $\chi^{i, \alpha}_R$ fields are equivalent to the anomaly of one
right-handed fermion in the fundamental representation of SU(4)$_{ETC}$. This
plays the role of a right-handed electroweak-singlet neutrino-type ETC
multiplet, so that, in conjunction with the fermion fields in
Eqs. (\ref{qlgen}), (\ref{urgen}), (\ref{drgen}), (\ref{elgen}), and
(\ref{ergen}), it renders the part of SU(4)$_{ETC}$ involving SM-nonsinglet
fermions vectorlike, so the $[{\rm SU}(4)_{ETC}]^3$ anomaly from these fermions
vanishes. The hypercolor sector is constructed so that its contribution to the
$[{\rm SU}(4)_{ETC}]^3$ anomaly also vanishes, so the entire $[{\rm
SU}(4)_{ETC}]^3$ anomaly is zero. The anomalies of the form
\beqs
& & 
[{\rm SU}(4)_{ETC}]^2 \, {\rm U}(1)_1, \quad 
[{\rm SU}(2)_L]^2 \, {\rm U}(1)_1, \quad
{\rm and} \ \ [{\rm U}(1)_1]^3 \cr\cr
& & 
\label{anom1}
\eeqs
cancel between (quarks plus techniquarks) and (leptons
plus technileptons). One also verifies that the following anomalies vanish: 
\beqs
& & 
[{\rm SU}(4)_{ETC}]^2 \, {\rm U}(1)_2, \quad 
[{\rm SU}(2)_L]^2 \, {\rm U}(1)_2, 
\cr\cr & & 
[{\rm SU}(3)_1]^3, \quad [{\rm SU}(3)_2]^3, \quad [{\rm SU}(3)_{MC}]^3 
\cr\cr & & 
[{\rm SU}(3)_1]^2 \, {\rm U}(1)_1, \quad [{\rm SU}(3)_1]^2 \, {\rm U}(1)_2, 
\cr\cr & & 
[{\rm SU}(3)_2]^2 \, {\rm U}(1)_1, \quad [{\rm SU}(3)_2]^2 \, {\rm U}(1)_2, 
\cr\cr & & 
[{\rm SU}(3)_{MC}]^2 \, {\rm U}(1)_1, \quad 
[{\rm SU}(3)_{MC}]^2 \, {\rm U}(1)_2, 
\cr\cr & & 
[{\rm SU}(2)_{HC}]^2 \, {\rm U}(1)_1, \quad 
[{\rm SU}(2)_{HC}]^2 \, {\rm U}(1)_2, 
\cr\cr & & 
[{\rm SU}(2)_{UC}]^2 \, {\rm U}(1)_1, \quad 
[{\rm SU}(2)_{UC}]^2 \, {\rm U}(1)_2, 
\cr\cr & & 
[{\rm U}(1)_2]^3, \quad [{\rm U}(1)_1]^2 \, {\rm U}(1)_2, \quad 
[{\rm U}(1)_2]^2 \, {\rm U}(1)_1  \ . 
\cr\cr & & 
\label{anom2}
\eeqs
Several of these anomalies vanish trivially, owing to the fact that the SM
fermions have zero U(1)$_2$ hypercharge and the modular construction of the
theory.  Within a semiclassical picture that incorporates gravity, one would
also require that the mixed gauge-gravitational anomalies ${\cal G}^2 \, {\rm
U}(1)_1$ and ${\cal G}^2 \, {\rm U}(1)_2$ vanish, where here ${\cal G}$ denotes
graviton.  This requirement is satisfied.

One also must check that there are no global Witten anomalies (associated with
the homotopy group $\pi_4({\rm SU}(2))={\mathbb Z}_2$).  This requires that the
number of chiral fermions transforming as doublets under each of the SU(2)
gauge group be even.  For the sector of SU(2)$_L$-nonsinglet fermions, we have
$(N_{gen}-1+N_{TC})(N_c+1)=16$ chiral doublets.  For the SU(2)$_{HC}$ sector we
have $N_{ETC}=4$ (holomorphic) chiral doublets.  Finally, for the SU(2)$_{UC}$
sector we have $4N_{UC}=8$ (holomorphic) chiral doublets.  Thus, the theory is
free of any global $\pi_4$ anomaly.

\subsection{Symmetry Breaking of SU(4)$_{ETC}$ to SU(2)$_{TC}$}

The model is constructed so that the breaking of the SU(4)$_{ETC}$ symmetry to
SU(3)$_{ETC}$ at a scale $\Lambda_1$, and then to the residual exact
SU(2)$_{TC}$ symmetry at a lower scale $\Lambda_2$ is primarily driven by the
HC gauge interaction, which is arranged to become strong at a scale
$\Lambda_{HC} \simeq \Lambda_1 \sim 10^3$ TeV.  The details of how an
SU(4)$_{ETC}$ theory can be broken to the residual exact technicolor subgroup
SU(2)$_{TC}$ were presented in our Ref. \cite{gen}, to which we refer the
reader.  Here we only briefly mention the main points.  One chooses the values
of the SU(4)$_{ETC}$ and SU(2)$_{HC}$ couplings at a high scale so that at the
scale $\Lambda_1$, the HC interaction is sufficiently stronger than the ETC
interaction that the most attractive channel involves HC-nonsinglet fermions
and is of the form (in the notation of
Eqs. (\ref{ngen2_psi})-(\ref{ngen2_zeta}))
\beq
(4,2) \times (6,2) \to (\bar 4,1) \ ,
\label{ngen2_mac1}
\eeq
with $\Delta C_2 = 5/2$ for SU(4)$_{ETC}$ and $\Delta C_2 = 3/2$ for
SU(2)$_{HC}$.  The associated condensate is
\beq
\langle \epsilon_{ijk\ell}\epsilon_{\alpha\beta} \,
\chi^{j,\alpha \ T}_R C \zeta^{k\ell,\beta}_R \rangle \ ,
\label{ngen2_mac1condensate}
\eeq
where $\epsilon_{ijk\ell}$ is the totally antisymmetric tensor density for
SU(4)$_{ETC}$.  This breaks ${\rm SU}(4)_{ETC}$ to ${\rm SU}(3)_{ETC}$ and is
invariant under SU(2)$_{HC}$.  With no loss of generality, we may define the
uncontracted SU(4)$_{ETC}$ index in Eq. (\ref{ngen2_mac1condensate}) to be
$i=1$. This condensate also breaks the global U(1)$_{HCg}$ symmetry, giving
rise to a Nambu-Goldstone boson (NGB). (Additional physics in a UV completion
could render this a PNGB.) In general, (P)NGB's have derivative
couplings and hence have interactions that vanish in the limit where the
center-of-mass energy $\sqrt{s}$ is much less than the scale of the symmetry
breaking, i.e., here, energies much smaller than $10^3$ TeV. Moreover, this 
particular NGB is a SM-singlet, which further suppresses its observable
effects.  We shall discuss the (pseudo)-Nambu-Goldstone bosons (PNGB's)
resulting from the technifermion condensates below. 

As the theory evolves to lower energy scales, the SU(3)$_{ETC}$ and
SU(2)$_{HC}$ gauge couplings continue to grow, and at the scale $\Lambda_2$,
the dominant SU(2)$_{HC}$ interaction, in conjunction with the additional
strong SU(3)$_{ETC}$ interaction, produces a condensate in the most attractive
channel, which is (in a notation analogous to Eq. (\ref{ngen2_mac1})) 
\beq
(3,2) \times (3,2) \to (\bar 3,1) \ .
\label{ngen2_mac2}
\eeq
This has $\Delta C_2 = 4/3$ for SU(3)$_{ETC}$ and $\Delta C_2 = 3/2$ for
SU(2)$_{HC}$. The condensation in this channel breaks SU(3)$_{ETC}$ to
SU(2)$_{TC}$ and is invariant under SU(2)$_{HC}$. The associated condensate is
\beq
\langle \epsilon_{ijk}\epsilon_{\alpha\beta} \,
\zeta^{1j,\alpha \ T}_R C \zeta^{1k,\beta}_R \rangle \ ,
\label{ngen2_mac2zetazeta}
\eeq
where $i,j,k \in \{2,3,4\}$.  With no loss of generality, we may choose $i=2$
as the breaking direction in SU(3)$_{ETC}$.  Another condensate that is
expected to form at a scale slightly below $\Lambda_2$ is 
\beq
\langle \epsilon_{\alpha\beta}
\chi^{1,\alpha \ T}_R C \zeta^{12,\beta}_R \rangle \ ,
\label{ngen2_mac3chizeta}
\eeq
which does not break any further gauge symmetries beyond those broken at the
scales $\Lambda_1$ and $\Lambda_2$.  The choices of $\Lambda_1 \sim 10^3$ TeV
and a somewhat smaller value of $\Lambda_2$ can yield reasonable values for the
masses of the quarks and charged leptons of the first two generations.  Details
on SM fermion mass generation in this theory are given in Ref. \cite{gen}.
Since the $b$ quark and $\tau$ lepton are SU(4)$_{ETC}$-singlets, they would
have to get their masses in a manner different from the quarks and charged
leptons of the first two generations.  SU(3)$_1$-instanton effects can provide
a way to produce $m_b$ (via a 't Hooft determinantal operator) \cite{hill}.
Further ingredients are required to account for $m_\tau$ and to obtain the sort
of low-scale seesaw mechanism that was developed in Ref. \cite{nt} to explain
light neutrino masses (in a full SU(5)$_{ETC}$ theory).  Although there are no
intrinsic mass terms in the high-scale Lagrangian for the (SM-singlet) fermions
(\ref{ngen2_psi})-(\ref{ngen2_zeta}), these fermions all gain dynamical masses
of order $\Lambda_1$ or $\Lambda_2$ as a result of the various condensates that
form, and hence are integrated out of the effective low-energy theory below
$\Lambda_2$.

\subsection{Sequence of Condensations Involving $G_{ASM}$}

The breaking of $G_{ASM}$ is envisioned to occur at a scale roughly of order 1
TeV. In order to analyze this breaking, we first note that at this scale the
operative gauge symmetry is the one given in Eq. (\ref{gtcasm}).  Following
usual TC2 practice, the model is arranged so that $\alpha_{c1}(\mu)$ is
considerably larger than $\alpha_{c2}(\mu)$ at this scale of about 1 TeV.  This
inequality in couplings can arise naturally, since the leading coefficient of
the SU(3)$_1$ beta function is larger than the corresponding leading
coefficient of the SU(3)$_2$ beta function, as a consequence of the fact that
the SU(3)$_1$ sector has fewer fermions than the SU(3)$_1$ sector.

With the fermion content as specified above, the SU(3)$_1$ sector has $2+3=5$
Dirac fermions while the SU(3)$_2$ sector has $8+3=11$ Dirac fermions, both
transforming according to the respective fundamental representations of these
two groups. Hence, $b_1 = 23/3$ for SU(3)$_1$ while $b_1=11/3$ for
SU(3)$_2$. As the energy scale $\mu$ decreases through a value denoted
$\Lambda_t$, the coupling $\alpha_{c1}(\mu)$ grows to be sufficiently large
that the SU(3)$_1$ interaction produces a condensate in the $3 \times \bar 3
\to 1$ channel, namely
\beq
\langle \bar t t \rangle = \langle \bar t_{a,L} t^{a}_R \rangle + h.c.
\label{ttbarcondensate}
\eeq
This channel has an attractiveness measure $\Delta C_2 = 8/3$ with respect to
the SU(3)$_3$ gauge interaction.  If one assumes a given value of
$\alpha_{c1}(\mu_h)$ at a high scale $\mu_h$, then a rough estimate of the
value of the condensation scale $\Lambda_t$ can be obtained by using
Eq. (\ref{mucrit}) in the Appendix.  Owing to the formation of the condensate
(\ref{ttbarcondensate}), the top quark picks up a dynamical mass, and, indeed,
this comprises the dominant part of the mass of the top quark.  As discussed
above, the U(1)$_1$ interaction is attractive in this channel and repulsive in
the $\bar b b$ channel.  The condensate (\ref{ttbarcondensate}) breaks part of
the global symmetry group (\ref{su31globalsym}), namely ${\rm SU}(2)_{(t,b)_L}
\otimes {\rm SU}(2)_{(t,b)_R}$, to its diagonal subgroup, SU(2)$_{(t,b)_V}$,
yielding three (pseudo)-Nambu-Goldstone bosons, $|\pi_t\rangle$.  TC2 models
require that these be PNGB's rather than strictly massless NGB's because there
are also three NGB's $|\pi_{TC}\rangle$ resulting from the formation of the
technifermion condensates, with the same SM quantum numbers, and only one set
is absorbed to form the longitudinal components of the $W^\pm$ and $Z$.  The
$|\pi_t\rangle$ and $|\pi_{TC}\rangle$ mix to form the states
\beqs
& & |\pi_{EW} \rangle = \cos\theta |\pi_{TC}\rangle + \sin\theta |\pi_t \rangle
\cr\cr
& & |\pi_T \rangle = -\sin\theta |\pi_{TC}\rangle + \cos\theta |\pi_t \rangle
\label{pimix}
\eeqs
where $\theta$ is a mixing angle.  This mixing is relatively small,
corresponding to the fact noted above that the $W$ and $Z$ masses arise
primarily from the technicolor sector. The $|\pi_{EW}\rangle$ are absorbed by
the $W$ and $Z$, while the three orthogonal pseudoscalars $|\pi_T\rangle$ are
known as top pions. Using a Gell-Mann-Oakes-Renner-type formula, one infers
that the top pions will have masses given by $m_{\pi_T}^2 \simeq
-f_t^{-2}m_{t,res} \langle \bar t t \rangle$, where $m_{t,res}$ denotes a
contribution to $m_t$ that is hard on the scale of $\Lambda_t$ \cite{tc2}. This
constitutes another necessary ingredient in a satisfactory UV completion of the
TC2 model. With $\Lambda_t$ and $f_t$ as determined, $m_{\pi_T} \sim O(10^2)$
GeV.

A channel of the same $3 \times \bar 3 \to 1$ type with respect to SU(3)$_1$
and thus also a most attractive channel with respect to this group, is one that
would break the SU(3)$_{MC}$ gauge symmetry, namely one that would produce the
condensates
\beq
\langle \bar \eta_{a,p,L} \xi^{a,\lambda}_R \rangle \ , \quad p=1,2,3 \ .  
\label{xieta}
\eeq
This condensate also breaks the global SU(3)$_{\eta_L}$ and U(1) symmetries in
Eq.  (\ref{su31globalsym}), giving rise to (P)NGB's and also producing
dynamical masses of order $\Lambda_t$ for the $\eta^{a}_{p,L}$ fields.  As
before, the (P)NGB's are SM-singlets and are derivatively coupled, so their
effects at scales far below 1 TeV are suppressed.  These effects merit further
study.  We assume that if the vacuum alignment is such that the condensate
(\ref{xieta}) does form, it does so at a scale somewhat below $\Lambda_t \simeq
\Lambda_{MC}$ and hence does not significantly weaken the effective
SU(3)$_{MC}$ interaction at the scale $\Lambda_{MC}$.  The formation of the
condensate (\ref{xieta}) does have a positive role, since if this did not
happen, then the resultant low-energy effective field theory operative below
$\Lambda_{MC}$ would contain three light Dirac color-triplet,
electroweak-singlet fermions constructed from the six chiral fermions
$\eta^a_{p,L}$ and $\eta^{a'}_{p,R}$, $p=1,2,3$.  Experimentally, such states
are excluded, with lower limits of order several hundred GeV, depending on
details of the signatures of the production and decays GeV \cite{pdg,ess}.

Written in vectorlike form, the SU(3)$_{MC}$ gauge interaction has three Dirac
fermions transforming as the fundamental representation of this group.  This
number is well below the estimated critical number $N_{f,cr} \simeq 12$ beyond
which the theory would evolve into the infrared without spontaneously breaking
chiral symmetry (see Appendix for further discussion).  It follows that, as the
reference scale $\mu$ decreases through a scale denoted $\Lambda_{MC}$, the
SU(3)$_{MC}$ interaction gets sufficiently strong to cause condensation in the
channel $3 \times \bar 3 \to 1$, which is the most attractive channel, with
condensate
\beq
\langle \bar \xi_{a',\lambda,L} \xi^{a,\lambda}_R \rangle + h.c. \ , 
\label{xixicondensate}
\eeq
breaking ${\rm SU}(3)_1 \otimes {\rm SU}(3)_2$ to the diagonal subgroup,
SU(3)$_c$.  This condensation channel has an attractiveness measure $\Delta
C_2=8/3$ with respect to the SU(3)$_{MC}$ gauge interaction.  The fermions
involved in this condensate get dynamical masses of order $\Lambda_{MC}$ and
the gauge bosons (often called colorons) in the coset $[{\rm SU}(3)_1 \otimes
{\rm SU}(3)_2]/{\rm SU}(3)_c$ gain masses 
$(g_{c1}^2+g_{c2}^2)^{1/2}\Lambda_{MC} \simeq g_{c1} \Lambda_{MC} \simeq
\Lambda_{MC}$ (where the running couplings $g_{c1}$ and $g_{c2}$ are evaluated
at $\Lambda_{MC}$). The model is arranged so that $\Lambda_{MC} \lsim
\Lambda_t$; this is necessary since if $\Lambda_{MC}$ were larger than
$\Lambda_t$, then the SU(3)$_1$ would already have broken to regular color
SU(3)$_c$, which has a considerably weaker coupling (given by Eq. (\ref{gcrel})
below) at the scale $\Lambda_{MC}$ and hence would not produce a $\langle \bar
t t \rangle$ condensate.  On the other hand, $\Lambda_{MC}$ should be high
enough so that (i) the colorons have sufficiently large masses to avoid
conflict with the lower bounds from experimental searches
\cite{pdg,hillsimmons,sekhar_pdg} and also (ii) high enough to avoid
transitional threshold effects in high-precision $Z$ decay data, which are
consistent with equal color SU(3)$_c$ couplings of gluons to the
third-generation $b$ quark as well as first- and second-generation quarks.
A choice that plausibly satisfies these requirements is $\Lambda_{MC} \sim 1$ 
TeV.  The resultant color SU(3)$_c$ coupling is given by
\beq
\frac{1}{\alpha_c(\mu)} = \frac{1}{\alpha_{c1}(\mu)}+
\frac{1}{\alpha_{c2}(\mu)} \ , 
\label{gcrel}
\eeq
so that, with $\alpha_{c1}(\mu) >> \alpha_{c2}(\mu)$, the value of
$\alpha_c(\mu)$ at and below $\Lambda_{MC}$ is set by the weaker coupling,
$\alpha_{c2}(\mu)$, and can thus agree with measured values of $\alpha_c$ for
$\mu \le m_Z$.  In the effective low-energy field theory operative at scales
below $\Lambda_{MC}$, we shall use the index $a$ for the resultant SU(3)$_c$
color symmetry. The SU(3)$_{MC}$-induced condensate (\ref{xixicondensate}) 
breaks a further part of the (abelian subgroup of the) global symmetry group 
(\ref{su31globalsym}).

A comment is in order here concerning the choice of the metacolor gauge group.
One might ask why we do not use SU(2)$_{MC}$ rather than SU(3)$_{MC}$.  To see
why, let us imagine replacing SU(3)$_{MC}$ with SU(2)$_{MC}$, and thus using 
$p=1,2$ rather than $p=1,2,3$ in Eqs. (\ref{etaL}) and (\ref{etaR}). Then
as the SU(2)$_{MC}$ interaction becomes strong enough to cause condensation, 
the channel $2 \times 2 \to 1$ leading to the condensate (\ref{xixicondensate})
would have attractiveness measure $\Delta C_2 = 3/2$.  However, in this
hypothetical case where the metacolor group is SU(2)$_{MC}$, there would be an 
even more attractive condensation channel than the one producing the desired
condensate (\ref{xixicondensate}), namely the channel producing the 
undesired condensate
\beq
\langle \epsilon_{abc}\epsilon_{\lambda\rho} \xi^{a,\lambda \ T}_R C 
\xi^{b,\rho}_R \rangle \ , 
\label{xixiundesired}
\eeq
where $a, \ b, \ c$ are SU(3)$_1$ gauge indices and $\lambda, \ \rho$ are
indices for the hypothetical SU(2)$_{MC}$ group.  This condensate is undesired 
because it would break SU(3)$_1$ to SU(2)$_1$, and this, in turn, would prevent
the theory from yielding the usual color SU(3)$_c$ group at lower scales. With
the hypothetical SU(2)$_{MC}$ gauge interaction, this undesired condensation
channel would have the same degree of attractiveness $\Delta C_2 = 3/2$ with 
respect to SU(2)$_{MC}$, but it would have an additional attraction due to the
SU(3)$_1$ interaction, since it involves the SU(3)$_1$ channel 
$(3 \times 3)_a \to \bar 3$, with $\Delta C_2 = 4/3$.  Hence, by usual MAC
arguments, if the metacolor group were to be taken to be SU(2)$_{MC}$ instead
of SU(3)$_{MC}$, the unwanted condensate (\ref{xixiundesired}) would form
first, as the theory evolved toward the infrared, rather than the desired
condensate (\ref{xixicondensate}).  The use of SU(3)$_{MC}$ avoids this and
guarantees that the MAC is the one that produces the condensate
(\ref{xixicondensate}).

We next proceed to analyze the condensate produced by the SU(2)$_{UC}$
interaction.  The SU(2)$_{UC}$ sector is asymptotically free and, when written
in vectorial form, contains two Dirac fermions transforming as the fundamental
representation. This is well below the estimated critical number, $N_{f,cr}
\simeq 8$ (see Appendix) beyond which this theory would evolve into the
infrared in a chirally symmetric manner.  Hence, as the scale $\mu$ decreases
from large values through a value $\Lambda_{UC} \simeq \Lambda_{MC}$, the
coupling $\alpha_{UC}(\mu)$ grows large enough to trigger condensation in the
channel $2 \times 2 \to 1$, with condensate
\beq
\langle \epsilon_{\hat\alpha \hat\beta} \Omega^{\hat{\beta} \ T}_R C 
\tilde \Omega^{\hat{\alpha}}_R\rangle + h.c. 
\label{OOcondensate}
\eeq
This breaks the ${\rm U}(1)_1 \otimes {\rm U}(1)_2$ symmetry to the diagonal
subgroup, U(1)$_Y$.  The fermions $\Omega^{\hat{\alpha}}_R$ and
$\tilde\Omega^{\hat{\beta}}_R$ involved in this condensate gain dynamical
masses of order $\Lambda_{UC}$.  The condensate (\ref{OOcondensate}) preserves
the global U(1)$_{UCg}$ symmetry, which is, indeed, the global limit of the
U(1)$_Y$ symmetry acting on the UC-nonsinglet fermions.  We denote the gauge
fields corresponding to U(1)$_1$ and U(1)$_2$ as $B_1$ and $B_2$ (suppressing
Lorentz indices).  The linear combination
\beq
B' = \frac{-g'_1B_1 + g'_2B_2}{\sqrt{g_1^{'2} + g_{'2}}}
\label{bprime}
\eeq
gains a mass given by 
\beq
m_{B'}^2 \simeq \frac{(g_1^{'2} + g_2^{'2}) y^2 \, \Lambda_{UC}^2}{4} \ . 
\label{mbprimesq}
\eeq
The field (\ref{bprime}) corresponds to the 
coset $[{\rm U}(1)_1 \otimes {\rm U}(1)_2]/{\rm U}(1)_Y$.  
(The $B'$ is often called by the generic name $Z'$ in the literature.) 
The orthogonal linear combination, 
\beq
B = \frac{g'_2B_1 + g'_1B_2}{\sqrt{g_1^{'2} + g_{'2}}}
\label{b}
\eeq
is the gauge boson corresponding to the usual weak hypercharge U(1)$_Y$ in the
SM gauge group $G_{SM}$, which is massless at this stage, with squared coupling
$\alpha'(\mu) = g'(\mu)^2/(4\pi)$ given by
\beq
\frac{1}{\alpha'(\mu)} = \frac{1}{\alpha'_1(\mu)}+
\frac{1}{\alpha'_2(\mu)} \ . 
\label{alphay}
\eeq

At scales $\mu$ below $\Lambda_{UC} \simeq \Lambda_{MC}$, the effective gauge
group resulting from $G_{ASM}$ that is operative is thus $G_{SM}$.  The
residual symmetry resulting from $G_{ETCA}$ includes exact SU(2)$_{TC}$,
SU(2)$_{HC}$, and SU(2)$_{UC}$.  (The symmetry SU(2)$_{MC}$ may be broken by
condensates of the form (\ref{xieta}).)  Without further ingredients, the
fermions $\omega_{p,L}$ and $\omega_{p,R}$ with $p=1,2$ would remain in this
low-energy theory, as a vectorial pair of SU(2)$_L$-singlets with nonzero weak
hypercharge.  This would be problematic, since, from the standard relation
$Q=T_{3L}+(Y/2)$, it follows that they are electrically charged. Such light,
charged leptons are excluded for values of $y \sim O(1)$, and $|y|$ cannot be
made small compared to unity, because this would reduce the $B'$ mass too much.
It is possible that both this problem and the problem of a nearby Landau
singularity could be cured by the physics that would involve the embedding of
the ${\rm U}(1)_1 \otimes {\rm U}(1)_2$ symmetry in non-Abelian symmetry
group(s) operative at higher energy scales.

\subsection{SU(2)$_{TC}$ Technicolor Sector}

The final stage of condensation and resultant electroweak symmetry breaking
occurs at the scale $\Lambda_{TC}$, where the SU(2)$_{TC}$ interaction becomes
sufficiently strongly coupled to produce technifermion condensates, breaking
${\rm SU}(2)_L \otimes {\rm U}(1)_Y$ to U(1)$_{em}$.  We have discussed this in
general above. The resultant $W$ mass is given by Eq. (\ref{mwsq}).  A value of
$\Lambda_{TC}$ consistent with $f_{TC} \simeq 120$ GeV is $\Lambda_{TC} \simeq
250$ GeV.  With an SU(2)$_{TC}$ gauge group and $2N_{TD}=2(N_c+1)=8$ Dirac
technifermions, the technicolor sector can plausibly produce technifermion
condensates $\langle \bar F F \rangle$ (where $F$ refers to the techni-up
quarks, techni-down quarks, technineutrinos, and techni-charged leptons) and
exhibit walking behavior associated with an approximate infrared fixed point of
the TC beta function \cite{wtc,chipt,lgt}.  As mentioned above, this walking
behavior plays an important role in enhancing the values of SM fermion masses
produced by ETC (thereby enabling the theory to use higher ETC scales
$\Lambda_j$ which reduce flavor-changing neutral-current effects) and in making
possible a reduction in TC corrections to $W$ and $Z$ propagators, as measured
by the $S$ parameter \cite{wtc,lgt,cs10,scalc}.  The $|\pi_{EW} \rangle$ of
Eq. (\ref{pimix}) are absorbed to become the longitudinal components of the
$W^\pm$ and $Z$. This relies in part on the property that $\Lambda_{TC} <<
\Lambda_{UC}$, so that the massive $Z$ boson is the same linear combination, in
terms of the U(1)$_Y$ gauge field and the neutral SU(2)$_L$ gauge field as in
the Standard Model, namely $Z = \cos\theta_W B - \sin\theta_W A_3$.  Precision
electroweak data constrain deviations from this form and hence set a lower
bound of the TeV order on $\Lambda_{UC}$.  From these constraints and related
constraints on new sources of custodial symmetry violation, it follows that
$m_Z$ is given by the SM formula $m_Z = m_W/\cos\theta_W$.  In order for the
technicolor sector to be in agreement with precision electroweak data, it is
necessary that the walking behavior reduce its contributions to the $S$
parameter substantially relative to the value for a QCD-like theory
\cite{scalc}.  It is also necessary that the masses of the
pseudo-Nambu-Goldstone bosons generated by the chiral symmetry breaking due to
the technifermion condensates should be sufficiently large to evade current
experimental limits.  The walking behavior has been shown to help in this
regard \cite{wtc}, but these PNGB's are still a significant concern in this 
type of ETC model.

\section{On the Role of SU(2)$_L$}

The group $G_{ASM}$ involves modifications of two of the three factor groups of
the Standard Model, leaving the SU(2)$_L$ unmodified.  It is natural to ask
what the consequences would be of following an analogous procedure with this
group, replacing it with the direct product ${\rm SU}(2)_{L,1} \otimes {\rm
SU}(2)_{L,2}$ in a symmetry group operative at a scale well above 1 TeV.  We
find that this does not produce the desired $\langle \bar t t \rangle$
condensate.  To see what would happen, let us first consider a model with the
high-scale symmetry group $G$, as before, but with $G_{ASM}$ replaced by
\beqs
G_{ASM2} & = & {\rm SU}(3)_c \otimes {\rm SU}(2)_{L,1} \otimes
{\rm SU}(2)_{L,2} \otimes {\rm U}(1)_Y \ . \cr\cr
& &
\label{gasms}
\eeqs
The SM-nonsinglet fermions are taken to be as follows, where the numbers in
parentheses are the dimensions of the representations of SU(4)$_{ETC}$ and the
three non-Abelian factor groups in G$_{ASM2}$, and the subscript is the weak
hypercharge (all of these fields are singlets under ${\rm SU}(2)_{HC} \otimes 
{\rm SU}(3)_{MC} \otimes {\rm SU}(2)_{UC}$): 
\beq
Q^{a,i,k'}_L  \ : \quad (4,3,1,2)_{1/3}
\label{qlgensu2}
\eeq
\beq
\tilde Q^{a,k}_L = {t^a \choose b^a_L}_L \ : \quad (1,3,2,1)_{1/3}
\label{tblgensu2}
\eeq
\beq
u^{a,i}_R \ : \quad (4,3,1,1)_{4/3}
\label{urgensu2}
\eeq
\beq
 \quad t^a_R \ : (1,3,1,1)_{4/3}
\label{trgensu2}
\eeq
\beq
d^{a,i}_R \ :  \quad (4,3,1,1)_{-2/3}
\label{drgensu2}
\eeq
\beq
 \quad b^a_R \ : \quad (1,3,1,1)_{-2/3}
\label{brgensu2}
\eeq
\beq
L^{k'}_L \ : \quad (4,1,1,2)_{-1}
\label{elgensu2}
\eeq
\beq
\tilde L^{k}_L = {\nu_\tau \choose \tau}_L \ : \quad (1,1,2,1)_{-1}
\label{taulgensu2}
\eeq
\beq
e^i_R \ ; \quad (4,1,1,1)_{-2}
\label{ergensu2}
\eeq
and
\beq
\tau_R \ : \quad (1,1,1,1)_{-2} \ . 
\label{taursu2}
\eeq
Here $a$, $i$, $k$, and $k'$ are the SU(3)$_c$, SU(4)$_{ETC}$, SU(2)$_1$, and
SU(2)$_2$ gauge indices. We shall, for the moment, leave the SM-singlet sector
unspecified; we comment on this later.  All of the non-Abelian gauge
interactions here are asymptotically free.  

We assume that the values of the corresponding gauge couplings are such that,
as the energy scale $\mu$ decreases, the first interaction to become
sufficiently strongly coupled to form a condensate is SU(2)$_{L,1}$.  Let us
denote the scale at which this occurs as $\Lambda_{L}$.  We note that the
SU(2)$_{L,1}$ sector, as specified so far, has $N_c+1=4$ chiral SM-nonsinglet
fermions, or equivalently, two Dirac fermions, so that it is well within the
phase where chiral symmetry breaking takes place in the infrared.  However, the
resultant condensates of third-generation fermions are not of the desired type.
The most attractive channel for the strongly coupled SU(2)$_{L,1}$ interaction
is $2 \times 2 \to 1$, and it produces several condensates in this channel.
The first of these is of the form $\langle \epsilon_{k \ell} \tilde Q^{a,k \
T}_L C \tilde Q^{b,\ell}_L \rangle$, where $\epsilon_{k \ell}$ is the
antisymmetric tensor density for SU(2)$_{L,1}$.  This is automatically
antisymmetric in SU(3)$_c$ indices and hence is proportional to
\beq 
\langle \epsilon_{abc} \epsilon_{k \ell } \tilde Q^{a,k \ T}_L C
\tilde Q^{b,\ell}_L \rangle = 
2\langle \epsilon_{abc} t^{a \ T}_L C b^b_L \rangle \ , 
\label{tbcondensate}
\eeq 
where $\epsilon_{abc}$ is the antisymmetric tensor density of SU(3)$_c$.  This
transforms as a $(3 \times 3)_{antisym} = \bar 3$ under SU(3)$_c$ and hence
breaks SU(3)$_c$ to a subgroup SU(2)$_c$. It also violates hypercharge and
electric charge.  The strong SU(2)$_1$ interaction would also produce the
condensate
\beq 
\langle \epsilon_{k \ell } \tilde Q^{a,k  \ T}_L C \tilde L^{\ell}_L
\rangle = \langle t^{a \ T}_L C \tau_L - b^{a \ T}_L C \nu_{\tau L} \rangle 
\ , 
\label{ttaucondensate}
\eeq
which also breaks SU(3)$_3$ to an SU(2)$_c$ subgroup and violates hypercharge
and electric charge.  In addition to breaking these gauge symmetries, the
condensate (\ref{tbcondensate}) breaks baryon number by $\Delta B = 2/3$,
while the condensate (\ref{ttaucondensate}) breaks $B$ by $\Delta B = 1/3$ and
lepton number $L$ by $\Delta L=1$.  For all of these reasons, one avoids trying
to construct a model with SU(2)$_L$ replaced by ${\rm SU}(2)_{L,1} \otimes
{\rm SU}(2)_{L,2}$.

\section{Conclusions}

In conclusion, we have carried out an exploratory construction and analysis of
an ultraviolet extension of a TC2 theory.  Let us assess the merits and
shortcomings of this model.  Among the partial successes is the fact that our
model includes self-consistent dynamical mechanisms that can account for (i)
the breaking of the ETC gauge symmetry down to the residual exact technicolor
symmetry, (ii) the breaking of ${\rm SU}(3)_1 \otimes {\rm SU}(3)_2$ to
SU(3)$_c$, and (iii) the breaking of ${\rm U}(1)_1 \otimes {\rm U}(1)_2$ to
U(1)$_Y$.  The model incorporates the characteristic feature of topcolor and
TC2 theories, that the large mass of the top quark arises from the formation of
a $\langle \bar t t \rangle$ condensate. Given that the resultant technicolor
sector plausibly has a large but slowly running gauge coupling governed by an
approximate infrared fixed point, the TC/ETC interactions can produce a
reasonable spectrum of masses for the first two generations of quarks and
charged leptons.

However, the model needs additional ingredients to be more realistic.  Although
we are able to embed two of the three SM fermion generations in an ETC
framework, the model cannot contain a full embedding of all three generations
in the ETC symmetry in the usual manner because the generational part of the
ETC symmetry is incompatible with the property that the first two generations
of SM fermions transform differently under $G_{ASM}$ from the third generation.
Consequently, this model only has mixing between the first two generations.  A
fully satisfactory explanation of the $\tau$ lepton mass and the masses and
mixings of neutrinos requires further ingredients, as does an explanation of
the contribution to $m_t$ that is hard on the scale $\Lambda_t$.  Moreover, the
model contains the non-asymptotically free ${\rm U}(1)_1 \otimes {\rm
U}(1)_2$ sector of topcolor and TC2 models. As usual in TC2 models, the
strength of the U(1)$_1$ interaction is bounded above both by its violation of
custodial symmetry and by the requirement that there not be any nearby Landau
pole.  One would hope in further work to embed this abelian product group in
some asymptotically free, non-Abelian symmetry group(s).  The condensates that
are formed by the HC and MC interactions entail the appearance of
(pseudo)-Nambu-Goldstone bosons, whose phenomenological consequences need
further study. Some of the new fermions with nonzero SM quantum numbers, in
particular, the color-singlet, SU(2)$_L$-singlet, charged $\omega$ and
$\tilde\omega$ fermions, do not pick up large dynamical masses and hence would
remain light, indicating the need for further ingredients to make the model
more realistic.  A general comment is that the model has a profusion of new
gauge interactions.  In an ultimate theory one would, of course, need to
explain the relative values of the many gauge couplings.  Finally, there are
the usual concerns with TC/ETC models such as ensuring that
pseudo-Nambu-Goldstone bosons that result from the technifermion condensates
are sufficiently heavy, and ensuring that technicolor contributions to the $S$
parameter are sufficiently small to agree with experimental constraints.

Nevertheless, given the interest in models of electroweak symmetry breaking
involving both technifermion and top-quark condensates, we believe that it is
valuable to carry out this type of exploratory investigation of possible
ultraviolet extensions. It is hoped that the present study will be of use in
elucidating some of the challenges that one faces in building ultraviolet
completions of TC2 theories.

\bigskip
\bigskip

This research was partially supported by the grant NSF-PHY-06-53342.

\section{Appendix}

In this Appendix we list some results used in the text.  The beta function of a
given factor group $G_j$ with running gauge coupling $g_j(\mu)$ is $\beta_j =
dg_j/dt$, where $t=\ln \mu$ and $\mu$ is the reference scale. In terms of
$\alpha_j$,
\beq
\frac{d\alpha_j}{dt} = - \frac{\alpha_j^2}{2\pi}
\left [(b_1)_j + \frac{(b_2)_j \alpha_j}{4\pi} + O(\alpha_j^2) \right ] \ , 
\label{beta}
\eeq
where we recall that the first two coefficients, $(b_1)_j$ and $(b_2)_j$, are
scheme-independent. The beta function with perturbatively calculated
coefficients is appropriate to describe the running of a gauge coupling that is
not too large, in the energy range where the corresponding gauge fields are
dynamical (i.e., above corresponding scales at which $G_j$ is broken).  For the
condensation channel (\ref{rchannel}), a solution of the Dyson-Schwinger
equation in the approximation of one-gauge boson exchange (often called the
ladder approximation) yields the condition for the critical coupling
$\alpha_{j,cr}$ (e.g, \cite{wtc})
\beq
\frac{3 \alpha_{j,cr} \Delta C_2}{2\pi} = 1 \ , 
\label{alfcrit}
\eeq
where $\Delta C_2$ was given in Eq. (\ref{deltac2}). Clearly, this is only a
rough estimate, in view of the strong-coupling nature of the physics.  

To investigate the infrared behavior of an asymptotically free $G_j$ gauge
interaction (i.e., one for which $(b_1)_j > 0$), one integrates the beta
function.  For sufficiently few fermions that are nonsinglets under $G_j$,
$(b_2)_j>0$ and the coupling will eventually exceed the critical coupling for
condensation in some most attractive channel.  As the number of fermions that
are nonsinglets under $G_j$ increases, the sign of $(b_2)_j$ eventually
reverses, and in this case, the (perturbative) two-loop beta function has a
zero away from the origin, at
\beq
\alpha_{j,IR} = -\frac{4\pi (b_1)_j}{(b_2)_j} \ . 
\label{alfir}
\eeq
Within the context of these approximations, if this $\alpha_{j,IR} >
\alpha_{j,cr}$, then as $\alpha_j$ exceeds $\alpha_{j,cr}$, there is
condensation in the most attractive channel.  In this case, the fermions
involved in this condensate gain dynamical masses and the evolution of the
theory further into the infrared is governed by a different beta function, so
that $\alpha_{j,IR}$ is only an approximate fixed point. If, on the other
hand, $\alpha_{j,IR} < \alpha_{j,cr}$ for any condensation channel, then the
$G_j$ sector evolves into the infrared without any fermion condensation or
associated spontaneous chiral symmetry breaking and $\alpha_{j,IR}$ is an
exact infrared fixed point.  In each of these cases, the behavior that occurs
as a particular $G_j$ gauge sector evolves toward the infrared can be changed
if other interactions break the $G_j$ symmetry, as happens, for example, with
SU(3)$_2$. 

Moreover, for an SU(2) gauge theory with chiral fermions transforming according
to the fundamental representation, the absence of a global Witten anomaly means
that there must be an even number of these fermions, so that one can always
rewrite the theory in a vectorial form with $N_f$ Dirac fermions.  Then the
above analysis yields the estimate $N_{f,cr} \simeq 8$ for the critical number
of such fermions below which the theory has S$\chi$SB in the infrared
\cite{chipt}.  For a vectorial (asymptotically free) SU(3) theory with $N_f$
Dirac fermions transforming according to the fundamental representation, the
same type of analysis yields the estimate $N_{f,cr} \simeq 12$.  Recent lattice
results for SU(3) are broadly consistent, to within the theoretical
uncertainties, with this estimate \cite{lgt}.

Integrating Eq. (\ref{beta}) and imposing the condition (\ref{alfcrit}) yields,
to leading order, the following rough estimate for the energy scale where this
condensation occurs: 
\beq
\mu_{c,j} \simeq \mu_h \ {\rm exp} \left [ -\frac{2\pi}{(b_1)_j}
\left ( \alpha_j(\mu_h)^{-1} - \frac{3\Delta C_2}{2\pi} \right ) \right ] \ , 
\label{mucrit}
\eeq
where $\mu_h$ is a high-energy reference scale.


\begin{thebibliography}{99}
\bibitem{ngen3}
%
To be precise, the LEP measurement of the invisible width of the $Z$ shows
that there are $N_{gen}=3$ SM fermion generations with associated
neutrinos that are light compared with $m_Z/2$.
\bibitem{csm}
%
Refs. \cite{ckm,kt} also considered a different type of SU(5)$_{ETC}$ model in
which $Q_L$, $u_R$, and $d_R$ transform as \underline{5}, \ \underline{5}, and
$\bar{\underline{5}}$, respectively. This type of ETC model is able to generate
quark mixing without the necessity for the above-mentioned hypercharge effects,
but was shown to be excluded by flavor-changing neutral-current constraints.
\bibitem{casimir}
%
The quadratic Casimir invariant $C_2(R)$ for the representation $R$ of a Lie
group $G$ is given by $\sum_{a=1}^{o(G)} \sum_{j=1}^{dim(R)} [D_R(T_a)]_{ij}
[D_R(T_a)]_{jk} = C_2(R)\delta_{ik}$, where $a,b$ are group indices, $o(G)$ is
the order of the group, $T_a$ are the generators of the associated Lie algebra,
and $D_R(T_a)$ is the matrix of $T_a$ in the representation $R$.
\end{thebibliography}
\end{document}